# ARTICLE

# Accelerating Kinetics with Time-reversal Path Sampling

Zhirong Liu*[a]

In comparison to numerous enhanced sampling methods for equilbrium thermodynamics, accelerating simulations for kinetics and nonequilibrium statistics are relatively rare and less effective. Here we derive a time-reversal path sampling (tRPS) method based on the time reversibility to accelerate simulations for determining the transition rates between free-ernegy basins. It converts the difficult up-hill path sampling into an easy down-hill problem. The method is easy to implement, i.e., forward and backward shooting simulations with opposite initial velocities are conducted from random initial conformations within a transition-state region until they reach the basin minima, which are then assembled to give the distribution of transition paths efficiently. The effects of tRPS are demonstrated by comparison with direct simulations on protein folding and unfolding, where tRPS is shown to give results consistent with direct simulations and increase the efficiency by up to five orders of magnitude. The approach is generally applicable to stochatic processes with microscopic reversibility, no mater whether the variables are continuous or discrete.

## I. Introduction

Thermodynamics and kinetics are two fundamental aspects of physical chemistry. In general, thermodynamics and equilibrium statistics are relatively easy to handle compared to kinetics and nonequilibrium statistics. Based on the ubiquitous Boltzmann distribution, various efficient enhanced sampling methods have been developed to greatly accelerate molecular simulations of equilibrium properties,[1] e.g., umbrella sampling,[2,3] histogram method,[4,5] temperature replica-exchange,[6-8] integrated tempering sampling[9] and metadynamics[10,11]. In recent years, machine learning techniques were also combined with conventional enhanced sampling to explore the vast conformational space of molecules.[12-16]

In contrast, there is less known profound principles for kinetics and nonequilibrium statistics. The Onsager reciprocal relations caught keenly the time reversal symmetry of the underlying microscopic dynamics to express the equality of certain ratios between different pairs of forces and flows.[17-19] The Jarzynski equality,[20] which revealed an unexpected connection between irreversible work and free energy difference, actually utilized the invariance of equilibrium distributions. Kinetics was widely described by a transition state theory,[21] which is based only on the information of potential energy surfaces and thus cannot provide an accurate transition rate. For molecular simulations, a direct simulation on transition processes to determine the transition rate is usually inefficient since the simulation trajectory spends most of the time wobbling and swaying in the vicinity of the reactant free energy minimum and the transition events to the product basin are extremely rare.[22] A major category of accelerating kinetics simulations were based on a description of path ensemble where the transition paths can be sampled purposefully.[23] Many simulation methods have been developed accordingly, e.g., transition path sampling (TPS)[24-26], transition interface sampling (TIS)[27,28], forward flux sampling (FFS)[29,30] and a Bayesian relation method[31-33]. The distribution of path ensemble is inherently related to the maximum entropy principle,[34] and thus machine learning has potential applications in path statistics and kinetics computation. Other approaches, such as Hyperdynamics (using a bias potential to upshift the free energy basins)[35-37], transition path theory[38], reactive flux (Bennet-Chandler estimation of the transmission coefficient to correct the transition state theory approximation)[39,40] and the aimless shooting algorithm[41], were also explored. Overall, accelerating kinetics simulations often relied on more complex assumptions than enhanced samplings of thermodynamics, and are usually less effective.

In this letter, we present a related approach to accelerate kinetics simulation for systems with free energy barrier. The approach is established based on the time reversal symmetry of the microscopic dynamics and the existence of equilibrium distribution, and is thus expected to be generally applicable. The obtained formula to calculate transition rate is bias-free and is easy to implement.

## II. Theory

The microscopic state of a system is described by a point in the phase space. Imagine that a phase space has many states (similar to the ensemble picture) that evolve and bifurcate (due to stochastics) over time, forming some infinitely long trajectories. To make practical statistics on trajectories, one should use some ways to cut the infinitely long trajectories into paths (segments of trajectories) whose length (duration) is finite. In the literature, there are two main schemes in cutting trajectories. One is to cut trajectories at fixed time so that the resulting paths have the same duration. This scheme was

[a.] *College of Chemistry and Molecular Engineering, and Beijing National Laboratory for Molecular Sciences (BNLMS), Peking University, Beijing 100871, China.*
\* Corresponding author. E-mail address: LiuZhiRong@pku.edu.cn.

adopted by TPS[24-26] and S-shooting[42]. Another is to cut trajectories with some fixed planes in the phase space with specified conformational features. For example, for a system with a double-well free energy profile, the basin minima can be chosen as the cutting points, and the trajectory was cut whenever it crosses the cutting points [Fig. 1(a)]. This scheme was adopted in methods such as FFS[29, 30], TIS[27, 28] and the Bayesian relation method[31-33], and is also adopted in this study.

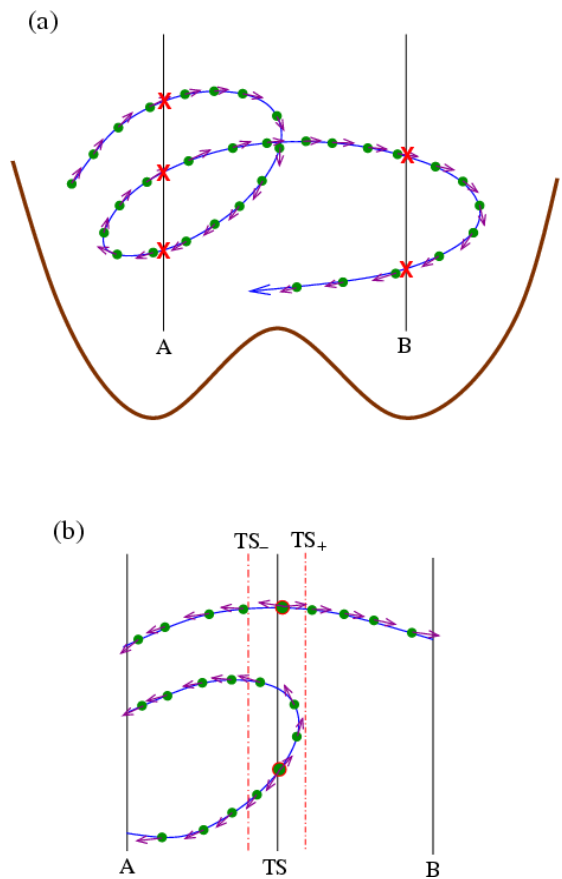

Figure 1. Schematics on accelerating kinetics with time reversibility. (a) The construction of path ensemble. A very long trajectory (blue line) is cut into some short paths (segments) by two cutting planes A and B located at the basin minima of the free energy profile (brown line). The cutting points are marked with red crossings. Green filled circles represent the conformations sampled on the trajectory with a fixed time step ($dt$) and violet arrows indicate their velocities. (b) Collection of paths crossing the transition state (TS). Starting from an initial conformation (bigger circles with red edge) (obeying equilibrium distribution) within the TS range $[TS_-, TS_+]$ (indicated by dashed-dotted lines), a forward and a backward simulations with opposite initial velocities are conducted until they reach any cutting planes (A or B), and, with the time reversal symmetry, they can be assembled to give a path passing the TS range.

The thermodynamic properties are determined by the state statistics, while the kinetic properties are embedded in the path statistics, i.e., the distribution of paths. To make the analysis more clear, the trajectories and paths are assumed to be composed of a sequence of conformations (plus velocities if necessary) sampled with a very small time step $dt$ [see Fig. 1(a)]. Then a path-$i$ ($\mathcal{P}_i$) is described in

$$\mathcal{P}_i = \{\mathcal{C}_{i,j}\}, \ j = 1, \ldots, L_i \quad (1)$$

where $\mathcal{C}_{i,j}$ is the $j$-th conformation of the path $\mathcal{P}_i$, and $L_i$ denotes the duration (in a unit of $dt$) of $\mathcal{P}_i$. Actually, the distribution of paths contains the information for state statistics. According to the basic postulates of statistical thermodynamics, the ensemble average (distribution) of a thermodynamic variable is equal to its time average. For a certain property $Q$ (which may present any conformation property, e.g., the fractional number of native contact for protein folding considered below, or the indicator to indicate whether the system falls in a certain free-energy basin A or B), its equilibrium distribution is given by

$$P(Q) \propto \sum_{i,j} H_Q(\mathcal{C}_{i,j}) \quad (2)$$

where the characteristic function $H_Q(\text{Conf}_{i,j}) = 1$ or 0 depending on whether $\mathcal{C}_{i,j}$ possesses a property $Q$. In other words, the number of conformations in a basin A in the ensemble (at any specified time) is equal to that of the conformations in A for all paths (at various time). When focusing on kinetics, one needs to calculate how many molecules transit from basin A to basin B in a certain time (Fig. 1). We use the basin minima A and B to cut the trajectory [Fig. 1(a)], and thus the paths can be classified into four types: A-A, A-B, B-A and B-B depending on the beginning and end points are cut by A or B. A-B and B-A are transition paths. Notably, each A-B path will contribute a transition event from A to B within $dt$, because each conformation (filled circles in Fig. 1) on a trajectory/path will run ahead (evolve) to occupy the position of the preceding one, regardless of the path duration and speeds, which is convenient for calculation. For the transition/reaction

$$A \xrightarrow{k} B, \quad (3)$$

the kinetic equation is

$$\frac{dN_A}{dt} = -kN_A \quad (4)$$

where $N_A$ is the number of molecules in basin A. Therefore, the transition rate coefficient $k$ is calculated as

$$k = -\frac{dN_A}{N_A dt} = \frac{1}{dt} \frac{\sum_i H_{A-B}(\mathcal{P}_i)}{\sum_{i,j} H_A(\mathcal{C}_{i,j})} \quad (5)$$

where the characteristic function $H_{A-B}(\mathcal{P}_i)$ indicates whether the path is classified into the type A-B, and $H_A(\mathcal{C}_{i,j})$ indicates whether the conformation falls in the basin A. The denominator in the above equation includes the contributions of paths A-A, A-B and B-A, but generally A-A is dominant and the other can be ignored to yield

$$k = \frac{1}{dt} \frac{\sum_i H_{A-B}(\mathcal{P}_i)}{\sum_i L_i H_{A-A}(\mathcal{P}_i)}. \quad (6)$$

Therefore, the kinetics can be readily obtained from path statistics. Eq. (6) is theoretically exact, so it is inherently equivalent to other path-based approaches with fixed-plane cuts[27-33]. But it is not very useful on its own since a direct sampling is usually inefficient, and thus further idea is needed as explained below.

In thermodynamics, sampling can be accelerated by applying bias or other means to enhance the sampling probability at certain regions of phase space. Similarly, methods can be developed to enhance the sampling of certain paths in kinetics simulations. Obviously, a direct simulation to generate trajectories/paths to calculate Eq. (6) is very inefficient since most produced paths belong to the type A-A and B-B but not transition paths A-B and B-A, and the resulting error of the numerator in Eq. (6) is large. Then it comes the main idea of this study [Fig. 1(b)], i.e., to enhance the sampling of A-B and B-A paths using the time reversal symmetry of the microscopic dynamics. Specifically, any A-B path will pass the transition state TS (strictly speaking, here we use it not for the genuine transition state, but just refer to a high free-energy range separating two basins, or, through which all transition paths have to pass), and is divided into an A-TS half-path and a TS-B half-path. The distribution of TS-B half-paths can be easily obtained by shooting simulation from initial states at TS (whose equilibrium distribution can be obtained by conventional enhanced sampling for equilibrium conformation statistics) and terminated at A or B (to give TS-A and TS-B half-paths). Although the A-TS half-paths is difficult to obtain in direct simulations from initial states at A, their population is exactly equal to that of the reverse TS-A ones due to the time reversal symmetry and can be obtained from the shooting simulations from TS. Therefore, the difficult up-hill path sampling is converted into an easy down-hill problem.

In actual simulations, the initial conformations are randomly chosen from an equilibrium distribution within a TS region $[TS_-, TS_+]$ which are determined by umbrella sampling in the examples below, and a forward and a backward shooting simulations with opposite initial velocities are conducted until they reach the basin minima A or B. Then the backward half-path is reversed and assembled with the forward one to give a path passing the TS range, which may be A-A (i.e., A-TS-A), A-B, B-A and B-B (i.e., B-TS-B) types. It is noted that such a path may contain more than one conformations in the TS region $[TS_-, TS_+]$ (the number of which is denoted as $n_{TS}$), which have the same probability to be chosen as initial state to generate the path. To avoid any double counting of the same path, the sampled paths by shooting simulations should be corrected by a weight $1/n_{TS}$ so as to be consistent with those in Eq. (6). In addition, it is noted that the denominator of Eqns. (5, 6) is actually the equilibrium conformation number in the basin A. Taking all these together, it yields

$$\begin{aligned} k &= \frac{1}{dt} \frac{N_{TS}}{N_A} \frac{\sum_i H_{A-B}(\mathcal{P}_i)}{N_{TS}} \\ &= \frac{1}{dt} \frac{N_{TS}}{N_A} \frac{\sum_{i \in \mathbb{S}} \frac{1}{n_{TS}} H_{A-B}(\mathcal{P}_i)}{\sum_{i \in \mathbb{S}} 1} \\ &= \frac{N_{TS}}{N_A} \frac{\sum_{i \in \mathbb{S}} \frac{1}{t_{TS}} H_{A-B}(\mathcal{P}_i)}{\sum_{i \in \mathbb{S}} 1} . \end{aligned} \quad (7)$$

where $\sum_{i \in \mathbb{S}}$ indicates a summation over paths assembled from shooting simulations, and $t_{TS} = n_{TS} dt$ is the duration of a path spent within the TS region. $N_{TS}/N_A$ is the population ratio of equilibrium conformations in the TS region and the basin A. Eq. (7) is the central result of this study, which indicates that the transition rate of kinetics can be obtained from an equilibrium result $N_{TS}/N_A$ and a shooting simulation, both of which are easy to implement. It is applicable to both continuous and discrete variables. We term the method time-reversal path sampling (tRPS).

It is noted that the forward/backward shooting moves with time-reversibility were widely employed in previous path samplings,[24-28, 31-33, 42] e.g., TPS,[24-26] TIS[27, 28] and the old-fashioned Bennett-Chandler approach[39]. It most cases, they were used as a means to perturb the old paths in order to provide trial paths for Monte-Carlo-like algorithm in constructing the path ensemble slice-by-slice, and the purpose is to calculate the correlation function or the conditional probability between adjacent slices used in rate formula. In comparison, here we utilized the time-reversibility to directly convert the difficult-to-calculate quantity (A-B paths) into easy-to-calculate quantity (TS-A and TS-B half-paths). We did not cast any sampled paths as did in the rejection/acceptance step in MC. We don't need to consider any other intermediate slices except the TS one. Visually speaking, in order to determine the height of the top of the steps, TIS and many closely related methods jump up stairs one by one, while tRPS directly jumps down from the top of the steps to the floor. See ESI for more details.

## III. Results

We test the method on the protein folding problem[43]. Although protein modelling has advanced rapidly over the past 50 years, a direct approach to protein folding kinetics is still challenging.[44] We consider a widely studied model protein, chymotrypsin inhibitor 2 (CI2) with 64 residues (PDB ID: 2CI2), which folds and unfolds in a simple two-state manner.[45] A coarse-grained Gō-like model was adopted to describe the conformational energetics of protein in the folding and unfolding processes,[46, 47] where the protein conformation is represented by the Cα coordinates of the amino acid residues (see Supporting Information). Molecular dynamics (MD) simulations were conducted to obtain the equilibrium conformation distribution and folding/unfolding rates.

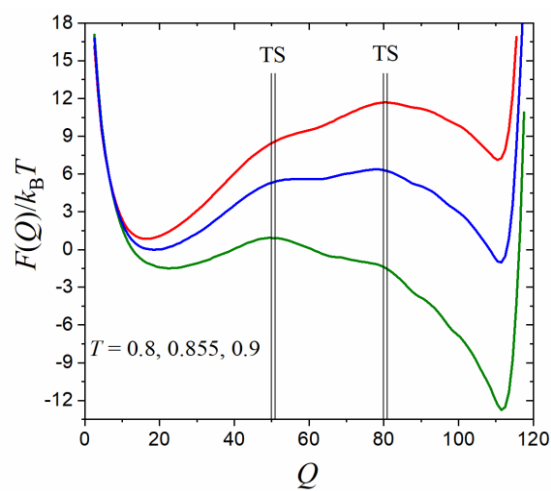

**Figure 2**. The free energy profiles of the protein CI2 as a function of the number of native contact ($Q$) at different reduced temperature (from top to bottom): $T = 0.9, 0.855, 0.8 \ \varepsilon/k_B$ (where $\varepsilon$ is the native contact energy strength). Two choices of transition state (TS) regions at around $Q$=50 and $Q = 80$ are indicated by black thin lines.

The obtained equilibrium free energy profiles of CI2 using the umbrella sampling technique are plotted in Fig. 2 as a function of the number of native contact ($Q$). The free energy profiles exhibit a typical double well form and are quite smooth, with one basin minimum at $Q \approx 20$ for unfolded states and another at $Q \approx 110$ for folded states. The folded states get more stable with decreasing the temperature. The midpoint temperature at which folded and unfolded states exhibit an equal stability was determined to be $T = 0.855$, where the free-energy barrier is about $6.4 k_B T$. We chose two schemes of TS regions in separating folded/unfolded basins, i.e., $Q_{TS} \in [50,51]$ and $Q_{TS} \in [80,81]$ (black thin lines in Fig. 2), to calculate the equilibrium population ratio $N_{TS}/N_A$ (here $A$ can represent unfolded or folded states) as well as preparing initial conformations within the TS region for shooting simulations as required by tRPS in Eq. (7).

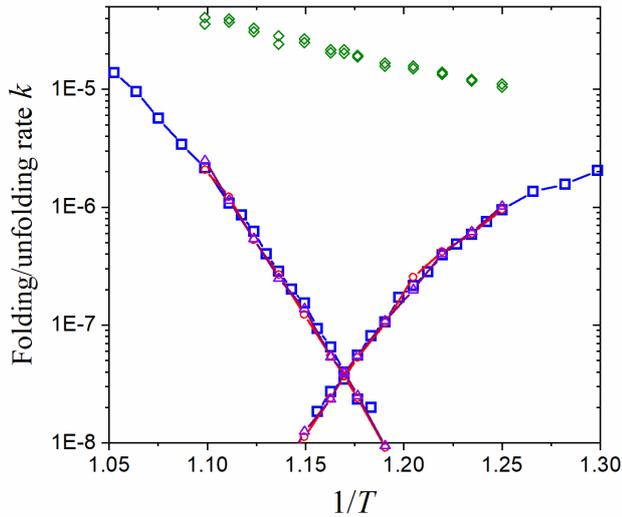

**Figure 3**. Accelerating folding/unfolding kinetics of CI2 with the tRPS method. Folding (right branches) and unfolding (left branches) rates were measured in a unit of $(\Delta t)^{-1}$ where $\Delta t$ is the MD time step. The temperature $T$ was measured in a unit of $\varepsilon/k_B$. The folding/unfolding rates of direct simulations were plotted in blue squares, while those obtained by tRPS [Eq. (7)] were shown in red circles (with $Q_{TS} \in [80,81]$) and violet triangles (with $Q_{TS} \in [50,51]$). The sampling rate of transition paths were plotted in scattering green diamond to demonstrate the acceleration effect. Each datapoint of direct simulations was averaged from about 400 folding/unfolding runs, while those of tRPS were each averaged from about 4000 paths.

The accelerating effect on the kinetics calculation using the tRPS method is demonstrated in Fig. 3 with a comparison to direct simulations. The logarithmic folding/unfolding rates form a typical V-shape (chevron plot in protein folding). The rates obtained from tRPS agree excellently with the direct simulations in a wide range over two orders of magnitude. It is noted that the choice of TS region, whether $Q_{TS} \in [50,51]$ or $Q_{TS} \in [80,81]$, does not affect the agreement since the only requirement for the TS region is that it separates the folded and unfolded basins. It is not required to be the true transition state. As a proof of the acceleration, the simulation time consumed in tRPS, e.g., the average time in obtaining a transition path in shooting simulation starting from TS region (green diamonds in Fig. 3), is much shorter than that in direct simulations (blue squares). Although the equilibrium property $N_{TS}/N_A$ is also required by Eq. (7), various efficient enhanced sampling methods have been developed previously in obtaining the equilibrium properties (among which we adopted the umbrella sampling here). In addition, the temperature-dependence of free-energy difference $[-k_B T \ln(N_{TS}/N_A)]$ is approximately linear (Fig. S1) and is thus relatively simple to determine.

Usual models of kinetics suggested that the system has to oscillate around the bottom of a basin for many times before it finally crosses the barrier to successfully transit into another basin. This is in line with the protein-folding example here: the averaged duration of A-A and B-B paths keeps roughly unchanged with increasing the temperature (Fig. S2), similar to the characteristics of a simple pendulum. To have one successful A-B or B-A transition, it has to oscillate $10^3 \sim 10^6$ times. The acceleration of tRPS originates from the fact that it does not need to spend a lot of time on the massive oscillations, but can directly sample the transition paths.

With tRPS, transition paths can be readily obtained for analyses (Fig. 4). The averaged duration of transition paths $\langle t_{TPath} \rangle$ is in an order of magnitude of $10^4$ steps and increases exponentially with decreasing temperature [Fig. 4(a)], much larger than that for A-A and B-B oscillation paths (in the order of magnitude of $10^2$ steps, see Fig. S2). The logarithmic $t_{TPath}$ roughly obeys a Gaussian distribution [Fig. 4(b)], similar to previous studies[22, 33]. Remarkably, a transition path usually has multiple times to cross the TS region, the number of which well obeys an exponential distribution, i.e., a memoryless distribution [Fig. 4(d)]. Consequently, the duration of a path spent within the TS region, $t_{TS}$, also obeys the exponential distribution [Fig. 4(c)]. At the midpoint temperature $T = 0.855$, the average crossing times is about 25, and an averaged $t_{TS}$ of about 540 steps. The large crossing times indicates that the conventional transition state theory would inevitable overestimate the transition rates since it assumes that a transition path crosses the TS region only once. Another discrepancy with the transition state theory is that the activation enthalpy of folding/unfolding (the minus slope of unfolding curve in Fig. 3 is about $62\varepsilon$ at midpoint temperature) is not equal to the enthalpy difference between the TS region and the unfolded/folded state (contributed by the $N_{TS}/N_A$ term in Eq. (7), which is about $72\,\varepsilon$ for unfolding) due to the contribution of the last term in Eq. (7).

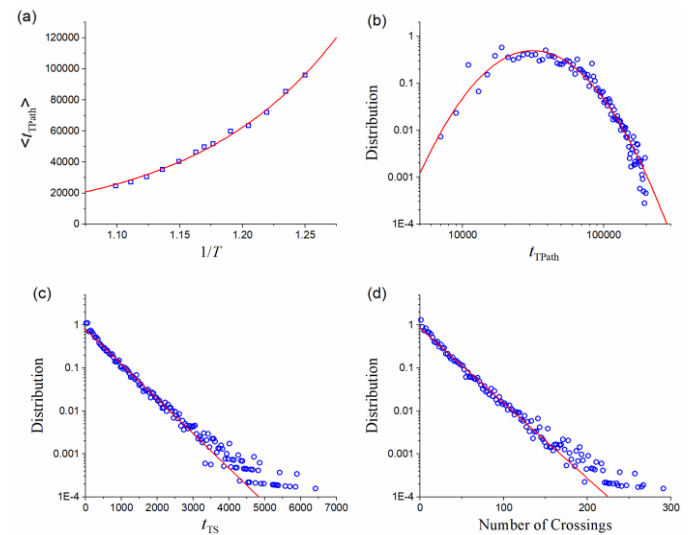

**Figure 4.** Properties of transition paths. (a) The averaged duration of transition paths ($t_{\mathrm{TPath}}$) as a function of $1/T$, which obeys an exponential law (solid line). (b, c, d) The distributions of $t_{\mathrm{TPath}}$ (b), the duration of a path spent within the TS region ($t_{\mathrm{TS}}$) (c), and the number of times a path crosses the TS region (d), at the midpoint temperature $T = 0.855$. Solid lines are quadratic fit in (b) and linear fit in (c,d).

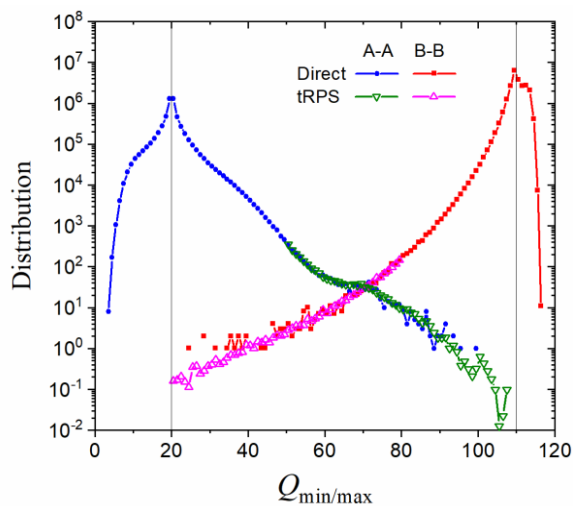

**Figure 5.** Distribution of minimal/maximal $Q$ for A-A and B-B paths at $T = 0.855$. $Q_A = 20$ and $Q_B = 110$ were used in cutting paths. Filled symbols represent datapoints from direct simulations, while open symbols for those from tRPS with $Q_{\mathrm{TS}} \in [50,51]$ (down triangles) and $Q_{\mathrm{TS}} \in [80,81]$ (up triangles).

The tRPS method can be combined with direct simulations to provide much more comprehensive results. For example, the minimal/maximal $Q$ value of a path can be adopted to measure how far it can go, and the resulting path population decreases exponentially with the distance between $Q_{\mathrm{min/max}}$ and the cutting point (Fig. 5). This makes the paths with distant $Q_{\mathrm{min/max}}$ hard to sample in direct simulations. The tRPS method, on the other hand, samples only the paths that cross the TS region and thus possess the capability to probe distant $Q_{\mathrm{min/max}}$. The patches they provided can be combined to give a smooth and complete distribution (Fig. 5).

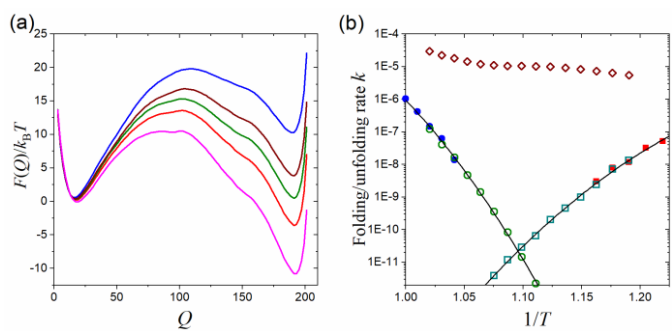

**Figure 6.** Equilibrium thermodynamics and kinetics of protein folding/unfolding for acylphosphatase. (a) The free energy profiles at different reduced temperature (from top to bottom): $T = 0.95, 0.925, 0.913, 0.9, 0.875$ $\varepsilon/k_B$. (b) Accelerating kinetics with tRPS. The temperature $T$ was measured in a unit of $\varepsilon/k_B$. The folding/unfolding rates of direct simulations were plotted in filled squares/circles, while those obtained by tRPS were shown in opened squares/circles (with $Q_{\mathrm{TS}} \in [100,102]$). The sampling rate of transition paths were plotted in scattering diamonds to demonstrate the acceleration effect.

Transition rates decay exponentially with the barrier height, but the duration of transition paths usually depends on the barrier in a much weaker logarithmic law.[22, 48] This makes tRPS even more powerful when the barrier is high. As a proof we apply tRPS on another protein, acylphosphatase (PDB ID: 1APS) with 98 residues, which was listed as a slow folding protein in a previous study[49]. The obtained free-energy profile is smooth, possessing a high barrier of about $15\ k_B T$ at a midpoint temperature of $T = 0.913$ [Fig. 6(a)]. The folding/unfolding are slow, being extremely difficult to determine with direct simulations. Therefore, we conducted direct simulations only in some feasible temperature range [filled circles/squares in Fig. 6(b)], and applied tRPS to complete the gaps (opened circles/squares). Results of direct simulation and tRPS in the overlapping area are well consistent with each other. The data combine to give a nice chevron plot. As a main expense of tRPS, the averaged duration [opened diamonds in Fig. 6(b)] of transition paths for acylphosphatase is similar in the order of magnitude with that for CI2. The increase in the efficiency of kinetics calculation by tRPS is up to five orders of magnitude around the midpoint temperature if not taking into account the expense of equilibrium calculations [for $N_{\mathrm{TS}}/N_A$ in Eq. (7)].

## IV. Discussion

The validity of tRPS relies on the time reversibility. Although it cannot be applied to irreversible systems as methods like FFS[29], it possesses the benefits of "shooting from the top"[33] to avoid possible inadequate choice of initial states in basin that are capable/incapable of crossing barrier. Although the TS region was not particularly optimized in our examples (Fig. S3), the resulting rates from tRPS are satisfactory (Figs. 3 and 6). Another underlying assumption of tRPS is the preequilibrium after a transition path, i.e., after a transition hit the cutting planes at the basin bottom, the system will pre-equilibrate within the basin but not cross the barrier back to the reactant basin soon. This can be tested by extending the shooting simulation after the path hits the cutting planes. Analyses on the examples of CI2 and acylphosphatase show that the error caused by the preequilibrium assumption is negligible (data not shown). In addition, Eq. (7) is beneficial in terms that it is less affected by possible complicated energy landscape around the bottom of basins. For the case of acylphosphatase, some abnormal high durations of transition paths were observed in shooting simulations at low temperatures (Fig. S4), likely caused by hidden traps within basins, but the resulting kinetic rates seem unaffected (Fig. 6).

Under harsh conditions such as hysteresis, the efficiency of tRPS may drop dramatically. This is a challenge that most path-based methods encountered. The procedure tRPS contains two parts: equilibrium sampling of TS region, and dynamic shooting simulations from the initial conformations obtained in TS region. In this study, the TS region was preassigned manually. In principle, self-adaptive TS region for purpose of efficiency optimization can be designed by allowing mutual interplays between two parts, i.e., let the shooting results also conversely affect the choice of TS region. Enhanced sampling methods and machine learning may play a role in it.

## Conclusions

In this letter, we have proposed a method to accelerate the simulations for determining the kinetics of systems. The approach was constructed based on the time reversibility of microscopic dynamics and is thus generally applicable. It is easy to implement, and can operate on both continuous and discrete variables. The method was tested on the folding/unfolding of two proteins with fast and slow kinetics. In areas where direct kinetics simulations can be readily performed, the accelerating method produced results fully consistent with direct simulations. In areas where direct simulations are inaccessible, the accelerating method provided reasonable results at little cost, with an increase of efficiency up to five orders of magnitude. The technique is easily applied to other kinds of calculations such as quantum dynamics and chemical reaction.

## Conflicts of interest

There are no conflicts to declare.

## Acknowledgements

This work was supported by the National Natural Science Foundation of China (grant 22273003). Part of the simulations was performed on the High-Performance Computing Platform of the Center for Life Science (Peking University).

SUPPORTING INFORMATION

# Accelerating Kinetics with Time-reversal Path Sampling

Zhirong Liu[*]

College of Chemistry and Molecular Engineering, and Beijing National Laboratory for Molecular Sciences (BNLMS), Peking University, Beijing 100871, China.

*E-mail: LiuZhiRong@pku.edu.cn

## MATERIALS AND METHODS

### Native-centric Gō-like model

We adopted a coarse-grained Gō-like model to describe the energetics of proteins in the folding and unfolding processes.[46, 47, 50] The protein conformation is represented by the Cα coordinates of the amino acid residues. Only native interactions are favorable in the model. The total potential energy of the model was given as

$$V_{\text{total}} = V_{\text{stretching}} + V_{\text{bending}} + V_{\text{torsion}} + V_{\text{non-bonded}}$$

$$= \sum_{\text{bonds } i} K_r \left(r_i - r_i^{(0)}\right)^2 + \sum_{\text{angles } i} K_\theta \left(\theta_i - \theta_i^{(0)}\right)^2$$

$$+ \sum_{\text{dihedrals } i} \left\{ K_\phi^1 \left[1 - \cos\left(\phi_i - \phi_i^{(0)}\right)\right] + K_\phi^3 \left[1 - \cos\left(3\phi_i - 3\phi_i^{(0)}\right)\right]\right\}$$

$$+ \sum_{i<j-3}^{\text{native}} U\left(r_{ij}; r_{ij}^{(0)}, \varepsilon\right) + \sum_{i<j-3}^{\text{nonnative}} \varepsilon \left(\frac{r_{\text{rep}}}{r_{ij}}\right)^{12}, \tag{S1}$$

where $r_i$, $\theta_i$, $\phi_i$ and $r_{ij}$ are the virtual bond length, bond angle, torsion angle and nonbonded spatial distance defined by Cα atom positions, respectively. $r_i^{(0)}$, $\theta_i^{(0)}$, $\phi_i^{(0)}$ and $r_{ij}^{(0)}$ are the corresponding native values available from the PDB structure. $U\left(r_{ij}; r_{ij}^{(0)}, \varepsilon\right)$ is a Lennard-Jones-like attraction potential with an extra solvation/desolvation barrier,[46, 50] which applies only for residue pairs in the native contact set. $U\left(r_{ij}; r_{ij}^{(0)}, \varepsilon\right)$ has a potential minimal $U = -\varepsilon$ at $r_{ij} = r_{ij}^{(0)}$. A pair of residues is defined to be in the native contact set if they are separated by at least three residues and a pair of their non-hydrogen atoms are less than 4.5 Å apart in the native PDB structure. $\varepsilon \left(\frac{r_{\text{rep}}}{r_{ij}}\right)^{12}$ is a nonnative repulsive term for any residue pairs being not in the native contact set. For CI2, the truncated form is composed of 64 residues,[45] and the PDB ID of its solved native structure is 2CI2. The resulting number of contacts in the native structure is $Q^{(N)} = 131$. A second protein considered in simulations, acylphosphatase (PDB ID: 1APS), contains 98 residues and $Q^{(N)} = 229$. Parameters of the model were set $r_{\text{rep}} = 4.0$ Å, $K_r = 100\varepsilon$, $K_\theta = 20\varepsilon$, $K_\phi^1 = \varepsilon$, $K_\phi^3 = 0.5\varepsilon$ with interaction strength $\varepsilon = 1.0\varepsilon_0$ (where $\varepsilon_0$ is a reference energy scale) as have been used previously.[47, 50]

### Molecular dynamics simulations

Simulations were performed in the form of Langevin dynamics:[47, 51]

$$m \frac{d}{dt} \mathbf{v}(t) = \mathbf{F}_{\text{conf}}(t) - m\gamma \mathbf{v}(t) + \mathbf{\eta}(t), \tag{S2}$$

where $m$, $\mathbf{v}$, $\mathbf{F}_{\text{conf}}$, $\gamma$ and $\mathbf{\eta}$ are mass, velocity, conformational force, friction (viscosity) constant and random force, respectively. The conformational force is equal to the negative gradient of the total potential energy $V_{\text{total}}$ in Eq. (S1). The time scale of simulations is controlled by the quantity $\tau = \sqrt{ma^2/\varepsilon_0}$, with the length scale $a = 4$ Å and a reference energy scale $\varepsilon_0 = 1$. The friction constant lies in the overdamped region with $\gamma = 1.0\tau^{-1}$. The molecular dynamics time step is set to be $\Delta t = 0.005\tau$. Simulation times in this study are presented in units of $\Delta t$. The temperature $T$ is given in a reduced form with a unit of $\varepsilon/k_B$. The coordinates are measured in units of Å.

The free-energy profiles as functions of number of native contact ($Q$) were obtained using the umbrella sampling method, and a smoothed version of $Q$ was adopted to facilitate the calculation of forces for umbrella bias potential.[37] The force constant of bias potential is $0.015\varepsilon_0$ for CI2 and $0.01\varepsilon_0$ for acylphosphatase. Typically, $8 \times 10^7$ time steps were used in simulations to obtained equilibrium conformation statistics for each bias potential.

Umbrella sampling was also used in preparing initial conformations within specified range of $Q$ for direct folding/unfolding simulations and the accelerating method of time-reversal path sampling (tRPS).

At each temperature, the folding/unfolding rate obtained by direct simulations was averaged from about 400 folding/unfolding runs, i.e., from 400 random initial conformations within the unfolded/folded basin. The rate obtained by tRPS was each averaged from about 4000 paths, i.e., a forward and a backward shooting simulations with opposite initial velocities were conducted from each of 4000 random initial conformations within a transition-state region.

**Theoretical comparison with TIS**

We first make a brief introduction on TIS and one of its variants, the replica exchange TIS (RETIS), based on the references[27, 28, 52]. The reaction rate was written as[27, 28, 52]

$$k_{AB} = f_A P(\lambda_B|\lambda_A) = f_A \prod_{i=0}^{n-1} P_A(\lambda_{i+1}|\lambda_i), \quad (S3)$$

where the phase space is divided by a set of $n$ interfaces (slices) $\{\lambda_A = \lambda_0, \ldots, \lambda_i, \ldots, \lambda_n = \lambda_B\}$. $f_A$ is the flux of paths through the initial interface $\lambda_A = \lambda_0$ per unit time. $P(\lambda_B|\lambda_A)$ is a conditional probability that a path starting from A, after having crossed $\lambda_A$, will cross the interface $\lambda_B$ before returning to $\lambda_A$. It can be conveniently factorized into the probabilities $P_A(\lambda_{i+1}|\lambda_i)$ that a path starting from A crosses the interface $\lambda_{i+1}$ after having crossed the interface $\lambda_i$ without returning to A first. Essential to construct the path ensembles (that crossed each interface $\lambda_i$) is to utilize MC-like moves. The most widely adopted MC move is the shooting move. In this move, a time slice of the last accepted path $\mathbf{x}^{(o)}$ is taken at random. Then, this point is modified, for instance, by changing the velocities of this point. Finally, this new point is used to shoot forward and backward in time using MD simulations in order to create a new (trial) path $\mathbf{x}^{(n)}$. The acceptance rule of the move for MC algorithm can be written as[28]

$$P_{\text{acc}}\left[\mathbf{x}^{(o)} \to \mathbf{x}^{(n)}\right] = \hat{h}(\mathbf{x}^{(n)}) \min\left[1, \frac{P[\mathbf{x}^{(n)}]}{P[\mathbf{x}^{(o)}]} \frac{P_{\text{gen}}[\mathbf{x}^{(n)} \to \mathbf{x}^{(o)}]}{P_{\text{gen}}[\mathbf{x}^{(o)} \to \mathbf{x}^{(n)}]}\right], \quad (S4)$$

where $\hat{h}(\mathbf{x}^{(n)})$ is 1 if the new path fulfills the path ensemble's condition (e.g., crossing $\lambda_i$), otherwise it is 0. $P_{\text{gen}}$ is the generation probability. If accepted, replace the old path with the new (trial) one; otherwise, cast the new (trial) path and keep the old one. Through this method, path ensembles can be constructed slice-by-slice, i.e., utilized the ensemble with paths crossing $\lambda_i$ to construct another ensemble with paths crossing $\lambda_{i+1}$, and calculate $P_A(\lambda_{i+1}|\lambda_i)$ in Eq. (S3).

From the above description, it can be known that the forward/backward shooting moves with time-reversibility were used in TIS/RETIS to perturb the old path into new (trial) path which was further fed into a Monte-Carlo-like algorithm of Eq. (S4). The new path inevitably shares some similarity with the old one. In other words, they are close in the path space. In some cases, the new (trial) path will be cast according to Eq. (S4).

In contrast, in our tRPS method, the time-reversibility was utilized to directly convert the difficult-to-calculate quantity (A-B paths, which contain A-TS or B-TS half-paths) into easy-to-calculate quantity (TS-A and TS-B half-paths). It is not necessary to consider a series of interfaces (slices). The forward/backward shooting was started from a conformation randomly chosen from the equilibrium distribution within the TS region $[TS_-, TS_+]$, but not from the previously obtained paths as done in TIS and RETIS. This is more efficient since equilibrium distribution is more easy to obtained (where enhanced sampling can be used), and the results paths in tRPS are uncorrelated, i.e., they are not necessarily close in the path space. In addition, no Monte-Carlo-like algorithm and acceptance/rejection step was used. Visually speaking, in order to determine the height of the top of the steps, TIS and many closely related methods jump up stairs one by one, while tRPS directly jumps down from the top of the steps.

For another approach, S-shooting[42], although a TS region is used without other slices, forward/backward shooting was still started from a conformation randomly chosen from the previously obtained paths as done in TIS and RETIS. The shooting points do not necessarily locate within the TS region. So it is different from tRPS.

# SUPPLEMENTARY FIGURES

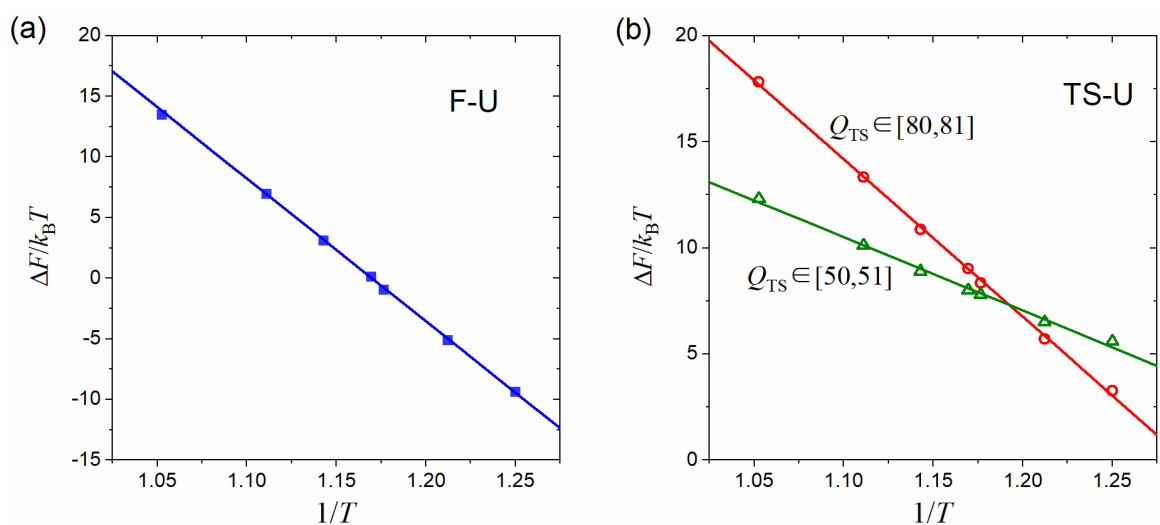

**Figure S1.** The temperature-dependence of free energy difference between: (a) folded and unfolded states [$\Delta F = -k_B T \ln(\frac{N_F}{N_U})$], (b) transition states and unfolded states [$\Delta F = -k_B T \ln(\frac{N_{TS}}{N_U})$]. The transition state region was defined as $Q_{TS} \in [50,51]$ (green) or $Q_{TS} \in [80,81]$ (red) in panel (b). Solid lines are linear fits to the data points.

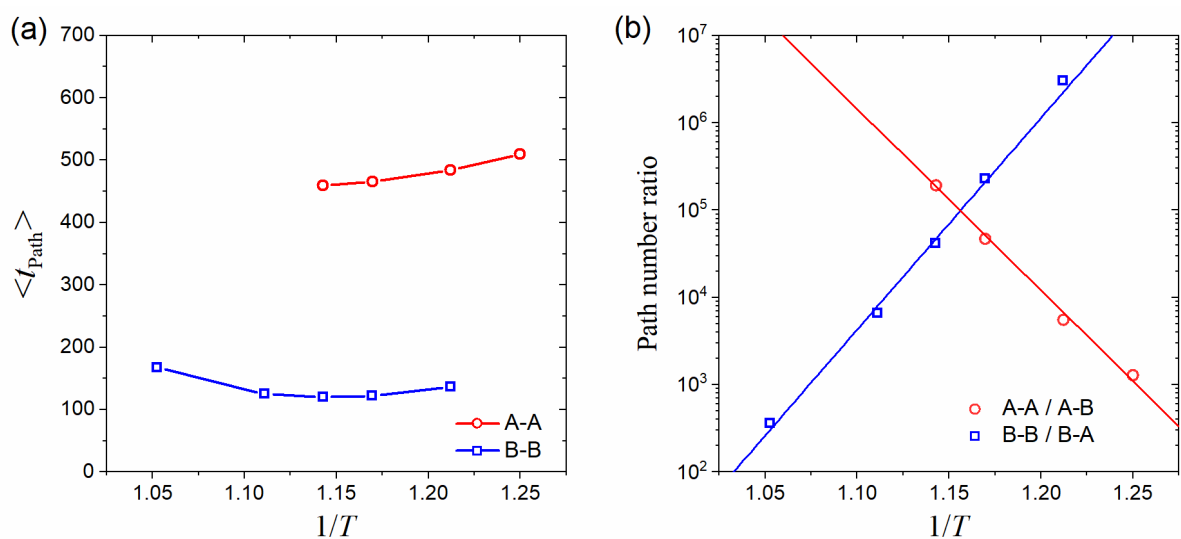

**Figure S2.** Some path properties determined from direct simulations. (a) The averaged duration of A-A and B-B paths ($t_{Path}$) as a function of $1/T$. (b) The path number ratio between A-A and A-B (or B-B and B-A) paths, where solid lines are linear fits. $Q_A = 50$ and $Q_B = 110$ were used in cutting paths.

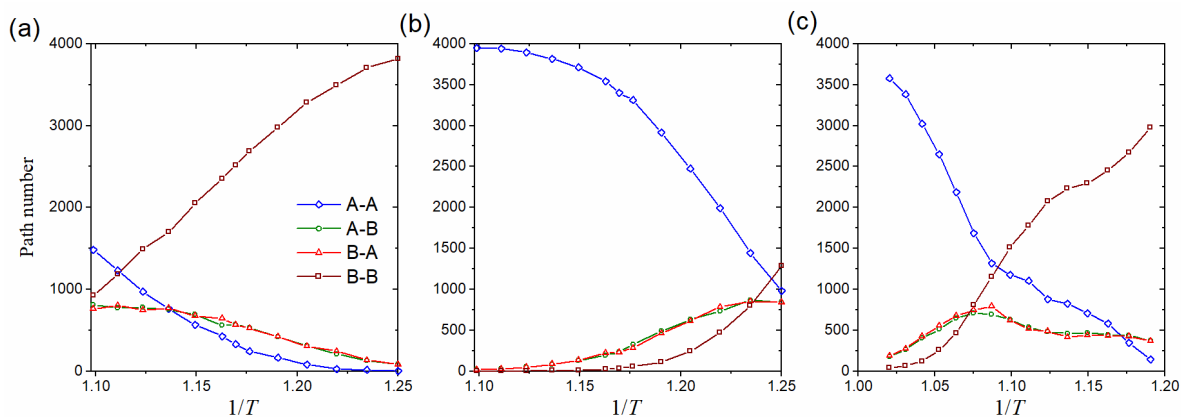

**Figure S3.** Path number sampled in tRPS for (a) CI2 with $Q_{TS} \in [80,81]$, (b) CI2 with $Q_{TS} \in [50,51]$, and (c) acylphosphatase with $Q_{TS} \in [100,102]$.

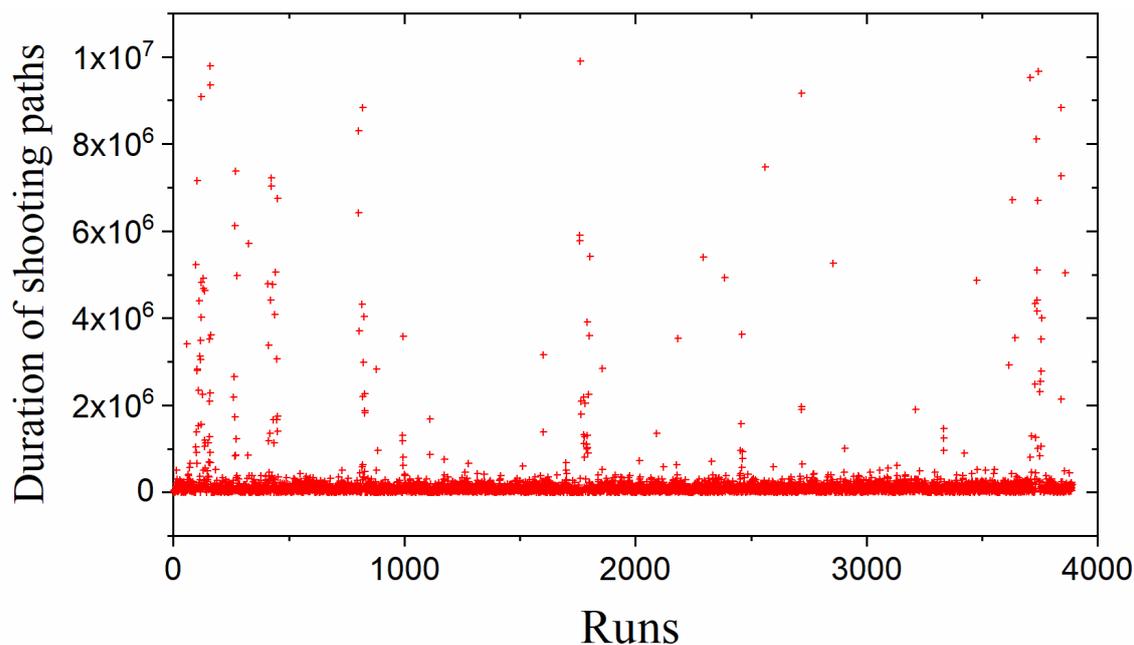

**Figure S4.** Duration of shooting paths from TS region for acylphosphatase at $T = 0.82$. The abnormal high values of duration are likely caused by traps inside the free-energy basins.